\documentclass[a4paper,twoside]{article}
\newcommand{\affil}[1]{$^{\rm #1}$}
\usepackage[authoryear]{natbib}
\bibpunct{(}{)}{;}{a}{}{,}
\usepackage{graphicx}
\date{} 
\newcommand{\Lya}{Ly$\alpha$~}
\title{\large\bf\flushleft Luminosity Bias (II): The
Cosmic Web of the First Stars}
\author{\parbox{\textwidth}{\flushleft
\vspace{-0.5cm}
{\it R.\ Barkana\affil{A,B}}\\
\vspace{0.4cm}
{\small \affil{A}\,Raymond and
Beverly Sackler School of Physics and Astronomy, Tel Aviv University,
Tel Aviv 69978, Israel}\\
{\small \affil{B}\,Email: barkana@wise.tau.ac.il}}}
\begin{document}
\twocolumn[
\begin{changemargin}{.8cm}{.5cm}
\begin{minipage}{.9\textwidth}
\vspace{-1cm}
\maketitle
%
%
\small{\bf Abstract:}

Understanding the formation and evolution of the first stars and
galaxies represents one of the most exciting frontiers in astronomy.
Since the universe was filled with neutral hydrogen at early times,
the most promising method for observing the epoch of the first stars
is using the prominent 21-cm spectral line of the hydrogen
atom. Current observational efforts are focused on the reionization
era (cosmic age $t\sim 500$ Myr), with earlier times considered much
more challenging. However, the next frontier of even earlier galaxy
formation ($t\sim 200$ Myr) is emerging as a promising observational
target. This is made possible by a recently noticed effect of a
significant relative velocity between the baryons and dark matter at
early times. The velocity difference suppresses star formation,
causing a unique form of early luminosity bias. The spatial variation
of this suppression enhances large-scale clustering and produces a
prominent cosmic web on 100 comoving Mpc scales in the 21-cm intensity
distribution. This structure makes it much more feasible for radio
astronomers to detect these early stars, and should drive a new focus
on this era, which is rich with little-explored astrophysics.


\medskip
\medskip
\end{minipage}
\end{changemargin}
]
\small

\section{Introduction}

\subsection{Cosmic reionization and the first stars}

\label{intro}

Galaxies around us have been mapped systematically out to a redshift
$z \sim 0.3$ by recent large surveys \citep{sdss,2df}. The observed
galaxy distribution shows a large-scale filament-dominated ``cosmic
web'' pattern that is reproduced by cosmological numerical simulations
\citep{mill}. This structure is well-understood theoretically
\citep{bond} as arising from the distribution of the primordial
density fluctuations, which drove hierarchical structure formation in
the early universe. Recent observations have been pushing a new
frontier of early cosmic epochs, with individual bright galaxies
detected reliably from as early as $z = 7.2$ \citep{z7p2}, which
corresponds to $t \sim 750$ Myr after the Big Bang. 

The refurbished Hubble Space Telescope [HST] and the Spitzer Space
Telescope have together provided increasingly detailed spectral
information on populations of $z \sim 6$ galaxies, making possible
robust determinations of their stellar masses, though substantial
uncertainties remain in the derived stellar ages and, thus, their
star-formation histories \citep{ono,schaerer,curtis}. The James Webb
Space Telescope [JWST], planned for launch in 2018, will have a 6.5
meter mirror and focus on infrared wavelengths \citep{gardner}. This
should allow it to reach much higher redshifts, detecting dwarf
galaxies and perhaps supernovae or gamma-ray burst afterglows.

A major goal of current and upcoming observations is to probe the era
of cosmic reionization. Ever since the discovery that the
intergalactic gas throughout the Universe is highly ionized
\citep{GP}, astronomers have been searching for the period when the
hydrogen was ionized for a second time after it became neutral at
cosmic recombination. This reionization is the most recent cosmic
phase transition, whereby the gas was heated and ionized throughout
the universe, affecting subsequent galaxy formation and potentially
detectable through its effect on a large range of observations.  While
the highest-redshift galaxies may be probing the late stages of
reionization, there is also a global constraint from the re-scattering
of some of the photons of the cosmic microwave background (CMB) by the
freshly created plasma. According to the WMAP and Planck CMB
satellites, the bulk of cosmic reionization occurred at $z \sim 11$
\citep{WMAP,Planck}.

While this observational progress has been remarkable, theory suggests
that a rich, varied history of early galaxy formation still remains to
be discovered. In particular, current and planned telescopes can
directly detect early galaxies only if they are particularly massive
or contain a bright, rare event. However, it is theoretically expected
that the bulk of the early stars formed in a large number of very
small galactic units, which would be difficult to observe
individually. Even the requirement of cosmic reionization likely
implies the existence of smaller galaxies than currently observed
[e.g., \citet{robertson}]. Going to even earlier times,
high-resolution numerical simulations suggest that the truly earliest
stars formed within $\sim 10^6 M_{\odot}$ dark matter halos
\citep{Abel,Bromm}.  Because this characteristic mass of early
galactic host halos is so small, constraining their abundance
observationally would probe primordial density fluctuations on $\sim
20$ kpc scales, an order of magnitude below current constraints. This
could lead to new limits on models with suppressed small-scale power
such as warm dark matter \citep{WDM}.

The first-star simulations mentioned above can only follow small
cosmic volumes, and thus begin to form stars much later than in the
real universe (since the simulations do not sample the over-dense
regions that are the sites of the earliest star formation), but
analytical methods show that the very first stars must have formed at
$z \sim 65$ (age $t \sim 35$ Myr) \citep{first,anastasia}.  The
formation of the very first star was a major milestone which ended the
dark ages of the universe -- the period after cosmic recombination,
when the universe was dark except for the fading glow of the CMB. It
also marked a transition point from the simplicity of the early
universe, which was homogeneous and isotropic except for small
fluctuations that can be described by linear perturbation theory. The
formation of the first stars initiated a new era of complexity and
feedback, whereby the nuclear and radiative processes within stars (on
the scale of $10^{11}$ cm) began to affect the global state of the
cosmic gas on scales larger by up to 15 orders of magnitude.

\subsection{21-cm cosmology}

\label{intro21}

The best hope of observing the bulk population of early stars is via
the cosmic radiation fields that they produced. The mean radiation
level traces the cosmic star formation rate, while spatial
fluctuations reflect the clustering of the underlying sources, and
thus the mass of their host halos. In particular, the hyperfine
spin-flip transition of neutral hydrogen (H~I) at a wavelength of 21
cm is potentially the most promising probe of the gas and stars at
early times. Observations of this line at a wavelength of $21\times
(1+z)$ cm can be used to slice the universe as a function of redshift
$z$ and obtain a three-dimensional map of the diffuse H~I distribution
within it \citep{Hogan}, in the previously unexplored era of redshifts
$\sim 7 - 200$.

Absorption or emission by the gas along a given line of sight changes
the 21-cm brightness temperature $T_b$, measured relative to the
temperature of the background source, which here is the CMB
\citep{Madau}. The observed $T_b$ is determined by the spin
temperature $T_s$, an effective temperature that describes the
relative abundance of hydrogen atoms in the excited hyperfine level
compared to the ground state. Primordial density inhomogeneities
imprinted a three-dimensional power spectrum of 21-cm intensity
fluctuations on scales down to $\sim 10$ kpc (all sizes comoving),
making it the richest data set on the sky \citep{Loeb04}. In
particular, since 21-cm measurements can in principle reach such small
scales and do it over a substantial redshift range, they can access
more independent modes, e.g., a factor of $\sim 10^9$ more than the
CMB, making 21-cm measurements potentially much more sensitive to
small levels of non-Gaussianity. The potential yield of 21-cm
observations is further enhanced by the expected anisotropy of the
21-cm power spectrum \citep{BL05a,Nusser,APindian,MeAP}.

The 21-cm signal vanished at redshifts above $z \sim 200$, when the
gas kinetic temperature, $T_k$, was close to the CMB temperature,
$T_\gamma$. Subsequently, the gas cooled adiabatically, faster than
the CMB, and atomic collisions kept the spin temperature $T_s$ of the
hyperfine level population below $T_\gamma$, so that the gas appeared
in 21-cm absorption \citep{Scott}. As the Hubble expansion continued
to rarefy the gas, radiative coupling of $T_s$ to $T_\gamma$ started
to dominate over collisional coupling of $T_s$ to $T_k$ and the 21-cm
signal began to diminish.

\begin{figure*}
\includegraphics[width=162mm]{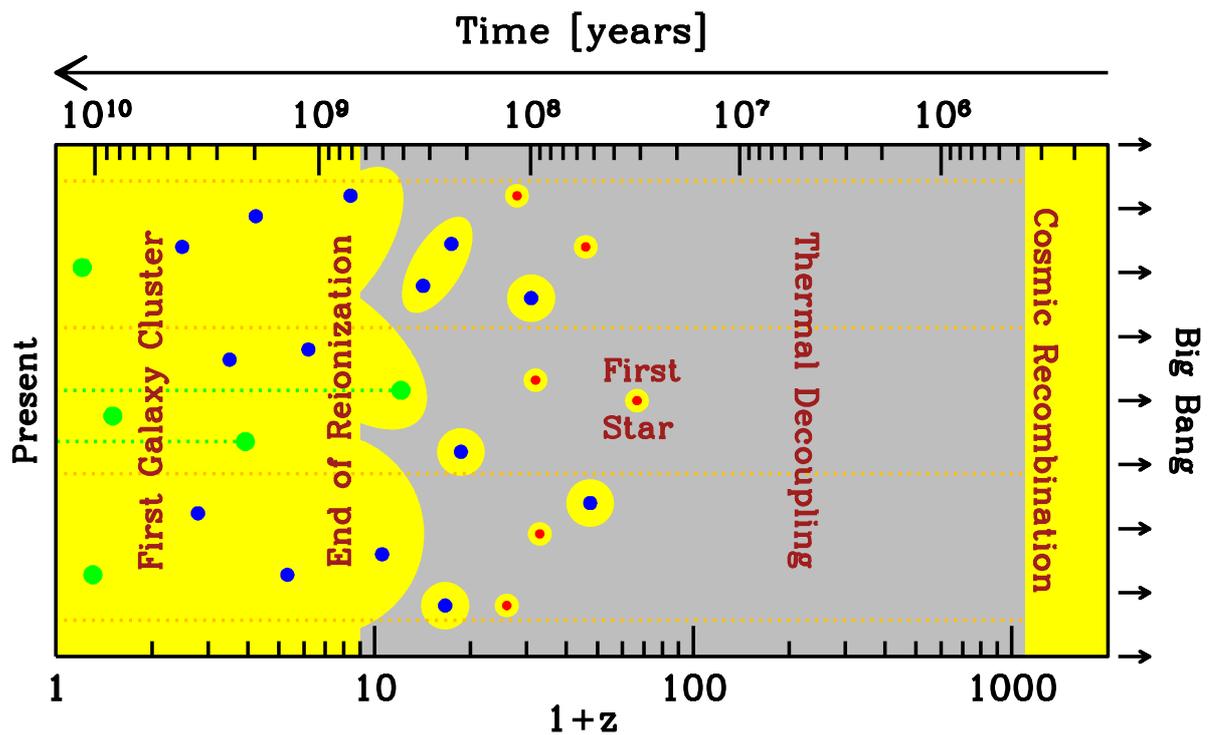}
\caption{Overview of cosmic history, with the age of the
universe shown on the top axis and the corresponding redshift on the
bottom axis. Yellow represents ionized hydrogen, and gray is
neutral. Observers probe the cosmic gas using the absorption of
background light (dotted lines) by atomic hydrogen. Stars formed in
halos whose typical size continually grew with time, going from the
first generation that formed through molecular-hydrogen cooling (red
dots), to the larger galaxies that formed through atomic cooling and
likely dominated cosmic reionization (blue dots), all the way to
galaxies as massive as the Milky Way, some of which host bright
quasars (green dots). From  \citet{BSc}.}
\label{f:history}
\end{figure*}

Once stars began to form, their radiation produced feedback on the
intergalactic medium (IGM) and on other newly-forming stars, and
substantially affected the 21-cm radiation. The first feedback came
from the ultraviolet (UV) photons produced by stars between the \Lya
and Lyman limit wavelengths. These photons propagated freely through
the universe, redshifted or scattered into the \Lya resonance, and
coupled $T_s$ to $T_k$ once again \citep{Madau} through the
Wouthuysen-Field \citep{Wout, Field} effect by which the two hyperfine
states are mixed through the absorption and re-emission of a \Lya
photon. Meanwhile, Lyman-Werner (LW) photons dissociated molecular
hydrogen and eventually ended the era of primordial star formation
driven by molecular cooling \citep{haiman}, leading to the dominance
of larger halos. X-ray photons also propagated far from the emitting
sources and began early on to heat the gas \citep{Madau}. Once $T_s$
grew larger than $T_\gamma$, the gas appeared in 21-cm emission over
the CMB level. Emission of UV photons above the Lyman limit by the
same galaxies initiated the process of cosmic reionization by creating
ionized bubbles in the neutral gas around these galaxies. See
Figure~\ref{f:history} for a brief summary of early cosmic history.

Several arrays of low-frequency radio telescopes are currently being
constructed in order to detect the 21-cm fluctuations from cosmic
reionization. Current efforts include the Murchison Wide-field Array
[MWA \citep{MWAref}], the Low Frequency Array [LOFAR
\citep{LOFARref}], the Giant Metrewave Radio Telescope [GMRT
\citep{GMRT}], and the Precision Array to Probe the Epoch of
Reionization [PAPER \citep{PAPER}], and early plans have been made for
a future Square Kilometer Array [SKA; e.g., \citet{F06}].  Although
the expected foregrounds (dominated by Galactic synchrotron) are much
brighter than the 21-cm signal, they are not expected to include sharp
spectral features. Thus, the prospects for extraction of the 21-cm
signal (and from it the reionization history) are quite promising,
using the 21-cm power spectrum \citep{Bowman,McQuinn,Sources} as well
as other statistics \citep{BiP,Fur04b,Ichikawa,Pan}. A different
approach is to measure the total sky spectrum and detect the global
reionization signal arising from the overall disappearance of atomic
hydrogen \citep{bowman2,PL,nono}.

\section{Large spatial fluctuations in galaxy numbers at 
high redshift}

A broad, common thread runs through much of the theoretical
development of early galaxy formation over the last decade: The
density of galaxies (or stars) varies spatially, with the fluctuations
becoming surprisingly large at high redshift, even on quite large
cosmological scales  \citep{BL04}. This can be understood from the
standard theory of galaxy biasing  \citep{ps74,k84,b86, ck89,
bond91,mw96} as due to the fact that the first galaxies represented
rare peaks in the cosmic density field. 

As an analogy, imagine searching on Earth for mountain peaks above
5000 meters. The 200 such peaks are not at all distributed uniformly
but instead are found in a few distinct clusters on top of large
mountain ranges. Similarly, in order to find the early galaxies, one
must first locate a region with a large-scale density enhancement, and
then galaxies will be found there in abundance. The density of stars
should thus show strongly biased (i.e., amplified) fluctuations on
large scales \citep{BL04}. These large-scale fluctuations, and their
effect on everything from feedback to observational predictions, had
been previously underestimated, in part because the limited range of
scales available to numerical simulations put this topic mostly out of
their reach.

\begin{figure}
\includegraphics[width=3.1in]{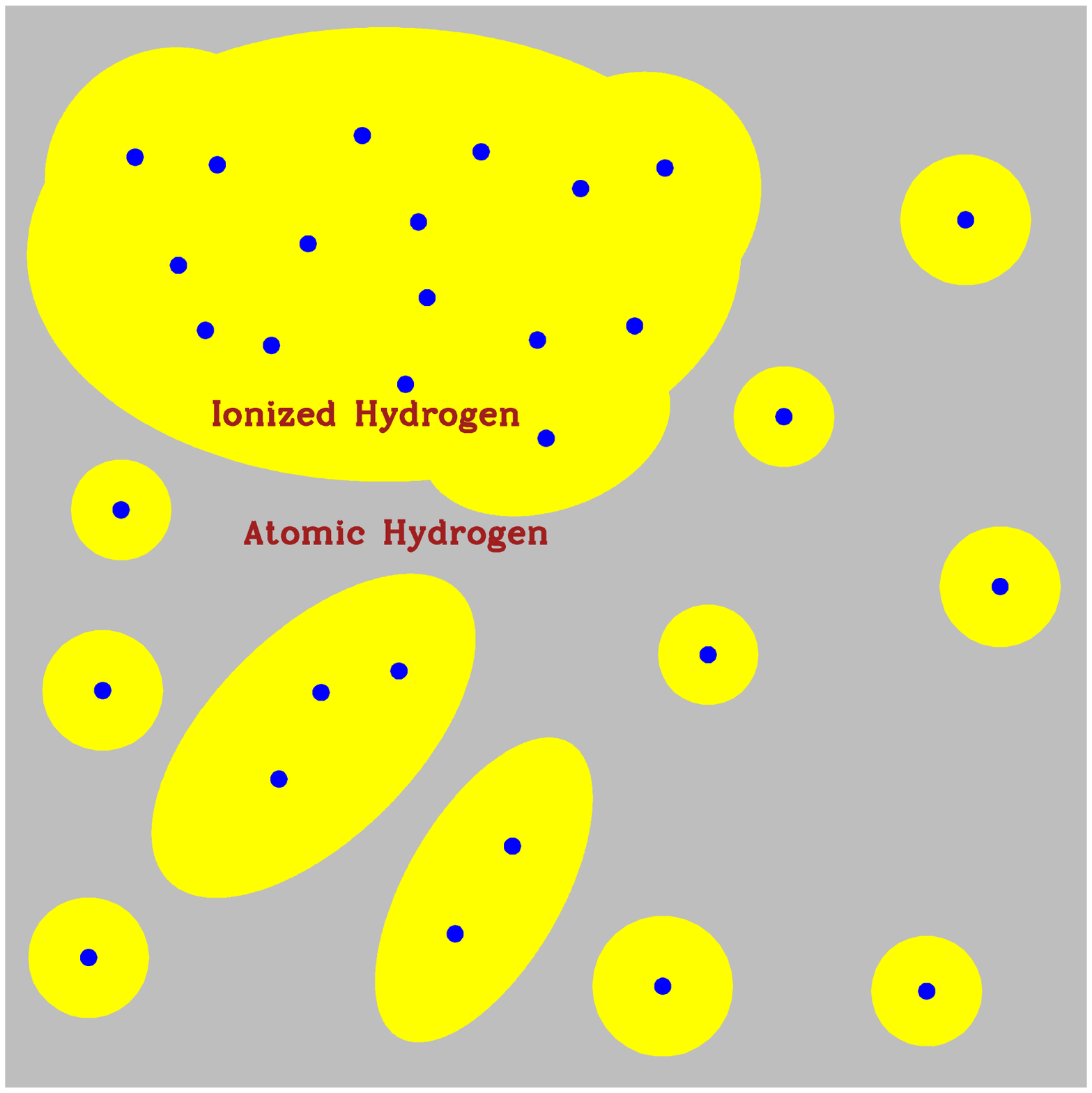}
\includegraphics[width=3in]{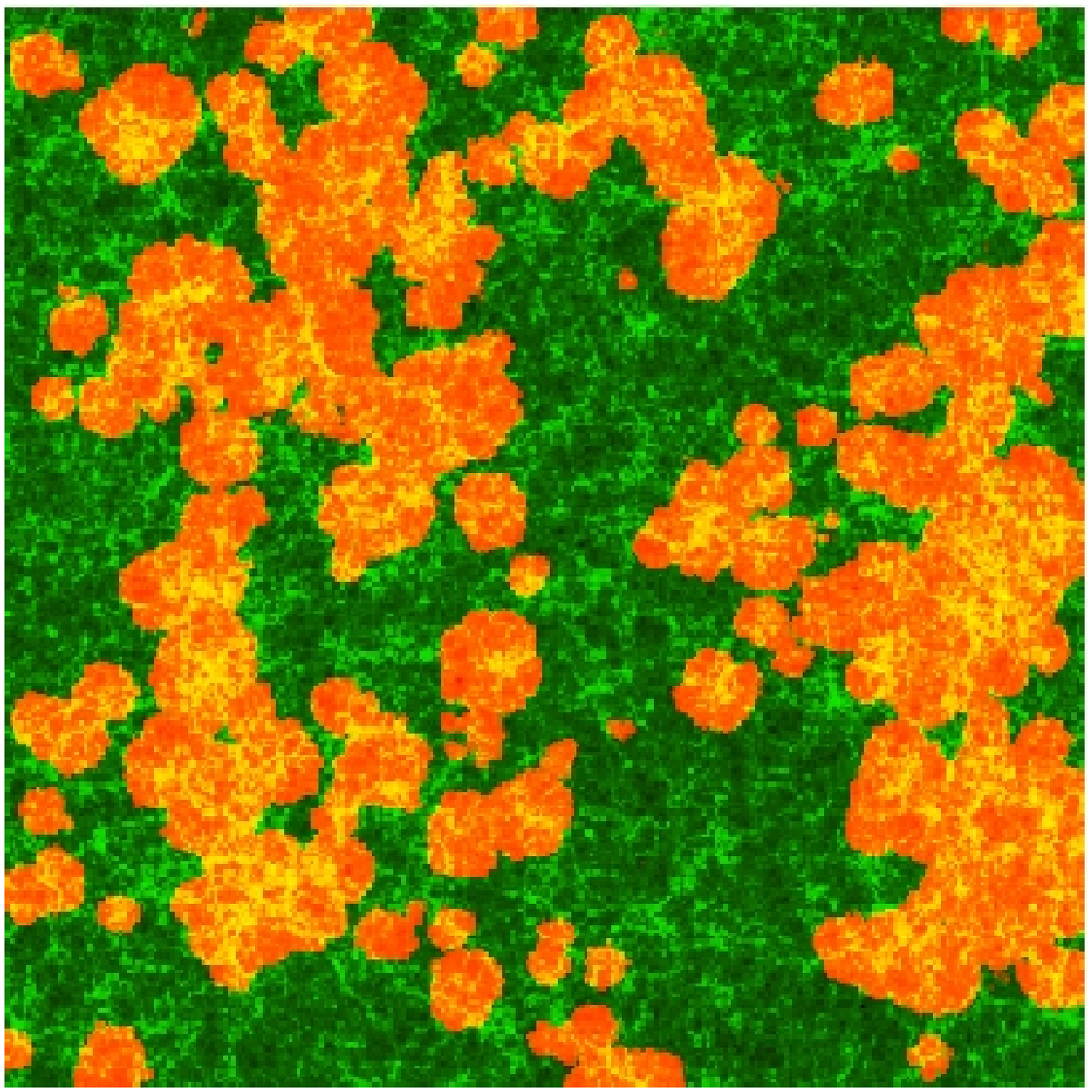}
\caption{During reionization, the ionized bubbles created by clustered
  groups of galaxies \citep{BL04} imprinted a signature in the power
  spectrum of 21-cm fluctuations \citep{fzh04}. The illustration (top
  panel, from \citet{BSc}) shows how regions with large-scale
  overdensities form large concentrations of galaxies (dots) whose
  ionizing photons produce large ionized bubbles. At the same time,
  other large regions have a low density of galaxies and are still
  mostly neutral. A similar pattern has been confirmed in large-scale
  numerical simulations of reionization (e.g., bottom panel, showing a
  two-dimensional slice from a 150 Mpc simulation box
  \citep{mellema}).  }
\label{f:reion}
\end{figure}

This idea first made a major impact on studies of cosmic reionization.
In \citet{BL04} we argued that the typical sizes of H~II bubbles
during reionization should be around 10 or 20 Mpc (see
Figure~\ref{f:reion}), while many numerical simulations of
reionization at the time followed a total box below this size. Further
analytical models \citep{fzh04} and large-scale numerical simulations
\citep{Ciardi,zahn,mellema,santos} have indeed demonstrated the
dominance of large bubbles due to large groups of strongly-clustered
galaxies. This has helped motivate the large number of observational
efforts currently underway, since large-scale fluctuations are easier
to detect (as they do not require high angular resolution).

The same idea of large-scale fluctuations in galaxy numbers soon found
an important application in a different regime. In \citet{BL05b} we
showed that fluctuations in the galaxy number density cause
fluctuations even in the intensity of long-range radiation, leading to
a new source of 21-cm fluctuations. Specifically, we considered the
background of \Lya radiation. As mentioned above, the spin temperature
of hydrogen atoms in the IGM can be coupled to the gas temperature
indirectly through the Wouthuysen-Field effect, which involves the
absorption of \Lya photons. While it was previously known
\citep{Madau, Miralda} that this \Lya coupling likely occurred in the
IGM due to \Lya photons emitted by early stars at $z \sim 20-30$, this
radiation background had been assumed to be uniform.  This intuition
was based on the fact that each atom sees \Lya radiation from sources
as far away as $\sim 300$ Mpc. However, we showed that relatively
large, potentially observable, 21-cm fluctuations are generated during
the era of initial \Lya coupling, for two reasons: galaxy fluctuations
are significant even on scales of order 100 Mpc, and also a
significant fraction of the \Lya flux received by each atom comes from
relatively nearby sources. Since a relatively small number of galaxies
contributed to the flux seen at any given point, we found that Poisson
fluctuations could be significant as well, producing correlated 21-cm
fluctuations.  If observed, we argued, this new predicted signal would
not only constitute the first detection of these early galaxies, but
the shape and amplitude of the resulting 21-cm power spectrum would
also probe their average properties.

This has led to a variety of follow-up work, including more precise
analyses of the atomic cascades of Lyman series photons
 \citep{Hirata,Jonathan06a} and a calculation of heating due to an
inhomogeneous X-rays background  \citep{Jonathan07} that followed a
similar approach and found somewhat larger resulting 21-cm
fluctuations. We have also predicted a significant boost in the 21-cm
power spectrum from \Lya fluctuations  \citep{NB08} due to the repeated
scattering of the photons from stars on their way to the hydrogen
atoms (see also \citet{Shapiro} and \citet{Semelin}).

\section{The new game in town: relative motion between the baryons 
and dark matter}

As noted above, current observational efforts in 21-cm cosmology (and
high-redshift astronomy more generally) are focused on the
reionization era (redshift $z \sim 10$), with earlier times considered
much more difficult to observe. This has begun to change as a result
of recent new predictions that suggest that the pre-reionization, $z
\sim 20$ era of even earlier galaxies is primed for observational
exploration.  This is made possible by a recently noticed effect on
early galaxy formation that had been previously neglected.

Up until recently, studies of early structure formation were based on
initial conditions from linear perturbation theory. However, there is
an important effect missing from this treatment \citep{TH10}. At early
times, the electrons in the ionized gas scattered strongly with the
then-energetic CMB photons, so that the baryons moved together with
the photons in a strongly-coupled fluid. On the other hand, the motion
of the dark matter was determined by gravity, as it did not otherwise
interact with the photons. Thus, the initial inhomogeneities in the
universe led to the gas and dark matter having different velocities.
When the gas recombined at $z \sim 1100$, it was moving relative to
the dark matter, with a relative velocity that varied spatially.  The
root-mean-square value at recombination was $\sim 30$ km/s, which is
supersonic (Mach number 5).

\begin{figure}
\includegraphics[width=3in]{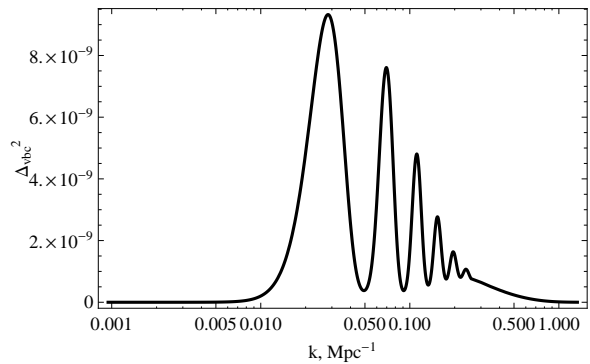}
\caption{The contribution of various scales to the mean squared
  velocity difference between the baryons and dark matter (at the same
  position) at recombination. The contribution per $\log k$ of
  fluctuations at wavenumber $k$ is shown vs.\ $k$. From \citet{TH10}.
}
\label{f:vbc}
\end{figure}

Figure~\ref{f:vbc} shows the contribution of fluctuations on various
scales to the variance of the velocity difference. This highlights two
important properties of this relative motion. First, there is no
contribution from small scales, so that the relative velocity is
uniform in patches up to a few Mpc in size; the velocity is generated
by larger-scale modes, up to $\sim 200$ Mpc in wavelength. The
uniformity on small scales is critical as it allows a separation of
scales between the spatial variation of the velocity (on large scales)
and galaxy formation (on small scales). Each individual high-redshift
mini-galaxy forms out of a small region ($\sim 20$ kpc for a $10^6
M_\odot$ halo) that can be accurately approximated as having a
uniform, local baryonic wind; the relative velocity is thus also
referred to as the ``streaming velocity''.  The second important
feature of Figure~\ref{f:vbc} is the strong baryon acoustic
oscillation (BAO) signature. Arising from the acoustic oscillations of
the photon-baryon fluid before recombination, this strong BAO
signature is a potential observational fingerprint of the effect of
this relative motion, as is further detailed below.

This relative motion effect is not in itself a surprise, but before
2010 it had not been noticed that this effect was dropped within the
standard approach. The standard initial conditions for both analytical
calculations and numerical simulations have been generated based on
linear perturbation theory, in which each $k$ mode evolves
independently. Indeed, the relative velocity is negligible if any
single scale is considered. However, it {\em is}\/ important as an
effect of large scales (which contribute to the velocity difference)
on small scales (which dominate early galaxy formation). Specifically,
the relative motion makes it harder for small-scale overdensities in
the dark matter to gravitationally accrete the streaming gas. Now,
observing such small-scale fluctuations directly would require far
higher resolution than is currently possible. Nonetheless, this effect
is immensely important because of the effect on star formation. Since
stellar radiation strongly affects 21-cm emission from the surrounding
IGM, 21-cm cosmology offers an indirect probe of the relative velocity
effect.

The first effect of the streaming velocity on halos to be analyzed was
the suppression of the abundance of halos \citep{TH10}. Since the
baryons do not follow the dark matter perturbations as closely as they
would without the velocity effect, linear fluctuations are suppressed
on small scales (where the gravitationally-induced velocities are
comparable to or smaller than the relative velocity). According to the
standard theoretical models for understanding the abundance of halos
as a function of mass \citep{ps74,bond91}, this should result in a
reduction of the number density of high-redshift halos of mass up to
$\sim 10^6 M_\odot$ \citep{TH10}, a mass range that is expected to
include most of the star-forming halos at early times.

\citet{Dalal} next pointed out a second effect, namely that separately
from the effect on the number of halos that form, the relative
velocity also suppresses the gas content of each halo that does form.
They also claimed that this second effect results in 2~mK, large-scale
21-cm fluctuations during \Lya coupling, with a power spectrum showing
a strong BAO signature due to the streaming velocity effect. These
conclusions were later shown to be qualitatively on the mark but
invalid quantitatively. In particular, we showed \citep{Us,anastasia}
that the gas-content effect is a minor one on star-forming halos, and
is mainly important for the lower-mass gas minihalos that do not form
stars. Also, a 2~mK signal would in any case be too weak to be of
major interest to the observers (see below).

Meanwhile, many groups began to run small-scale numerical simulations
that follow individual collapsing halos subject to the streaming
velocity
\citep{Maio:2011,Stacy:2011,Greif:2011,mcquinn12,mcquinn12b,naoz,naoz2}.
In particular, the simulations of \citet{Stacy:2011} and
\citet{Greif:2011} indicated the presence of a third effect, i.e.,
that the relative velocity substantially increases the minimum halo
mass for which stars can form from gas that cools via molecular
hydrogen cooling. We found \citep{anastasia} a fit to these simulation
results, and thus developed a general model that includes the effect
of density as well as all three effects of the velocity on star
formation.

\begin{figure*}[]
\includegraphics[width=162mm]{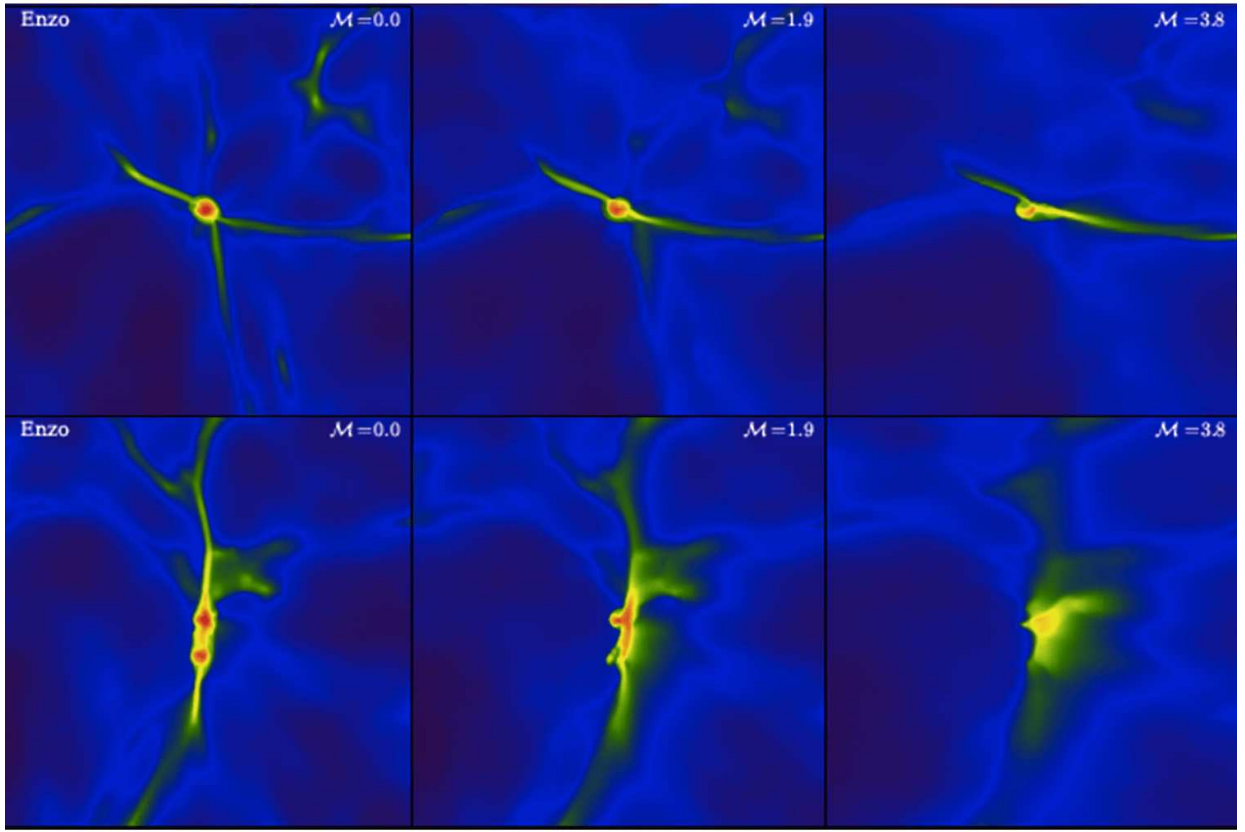}
\caption{Effect of relative velocity on individual halos,
from numerical simulations (including gravity and hydrodynamics).  The
colors indicate the gas density, which ranges from $10^{-26} {\rm
g/cm}^3$ (blue) to $10^{-23} {\rm g/cm}^3$ (red). Two halos are shown
at $z=20$, with a total halo mass of $2 \times 10^6 M_\odot$ (top) or
$8 \times 10^5 M_\odot$ (bottom). Panels show the result for gas
initially moving to the right with a relative velocity of 0 (left), 1
(middle), or 2 (right) in units of the root-mean-square value of the
relative velocity at $z=20$. $\mathcal{M}$ indicates the corresponding
Mach number at $z=20$. From  \citet{mcquinn12}.}
\label{f:numsim}
\end{figure*}

Figure~\ref{f:numsim} illustrates some of the results of the numerical
simulation studies of the effect of the streaming velocity on galaxy
formation. As expected, a larger velocity suppresses gas accretion
more strongly, in particular reducing the amount of dense gas at the
centers of halos. But beyond just this general trend, the relative
velocity effect gives rise to very interesting dynamics on small
scales. It disrupts gas accretion in an asymmetric way, so that
filaments of accreting gas are disrupted more easily if they are
perpendicular to the local wind direction. In addition, halos that
form in regions of relatively high velocity develop supersonic wakes
as they move through the wind.

\section{Detecting the First Stars at Redshift 20}

While numerical simulations are the best, most accurate method for
studying the small-scale effects of the relative velocity, they are
unable to simultaneously cover large volumes. Simulations that
successfully resolve the tiny mini-galaxies that dominated star
formation at early times are limited to $\sim 1$~Mpc volumes, and
cannot explore the large cosmological scales that might be accessible
to 21-cm observations (which are currently limited to low resolution).

On the other hand, analytical calculations are limited to linear (plus
sometimes weakly non-linear) scales, and thus cannot directly probe
the non-linear astrophysics of halo and star formation. Even if the
results of simulations are incorporated within them, analytical
approaches assume small fluctuations and linear bias (i.e., that the
distribution of stars is a proportionally amplified version of that of
the underlying density), assumptions that break down in the current
context, where the stellar density varies by orders of magnitude on
scales of a few Mpc. Even on 100 Mpc scales, fluctuations in the gas
temperature are as large as order unity (see below). There are also
additional non-linear relationships in this problem such as the
dependence of 21-cm temperature on gas temperature. Thus, linear,
analytical calculations can only yield rough estimates, even for the
large-scale fluctuations.

Thus, the best way to generate observable predictions from this era of
early galaxies is with a hybrid method that combines linear theory on
large scales with the results of numerical simulations on small
scales. We recently developed such a hybrid method and used it to
produce the first realistic, three-dimensional images of the expected
large-scale distribution of the first stars and the resulting 21-cm
emission \citep{nature}. In our approach we built upon previous hybrid
methods used for high-redshift galaxy formation
\citep{TH10,Dalal,21cmfast}. We first used the known statistical
properties of the initial density and velocity perturbations to
generate a realistic sample universe on large, linear scales. Then, we
calculated the stellar content of each pixel on our grid using the
overall model that we had developed \citep{anastasia} to describe the
streaming velocity effect on galaxy formation; this includes various
analytical models as well as fits to the results of the small-scale
numerical simulations.

\begin{figure}[]
\includegraphics[width=3.3in]{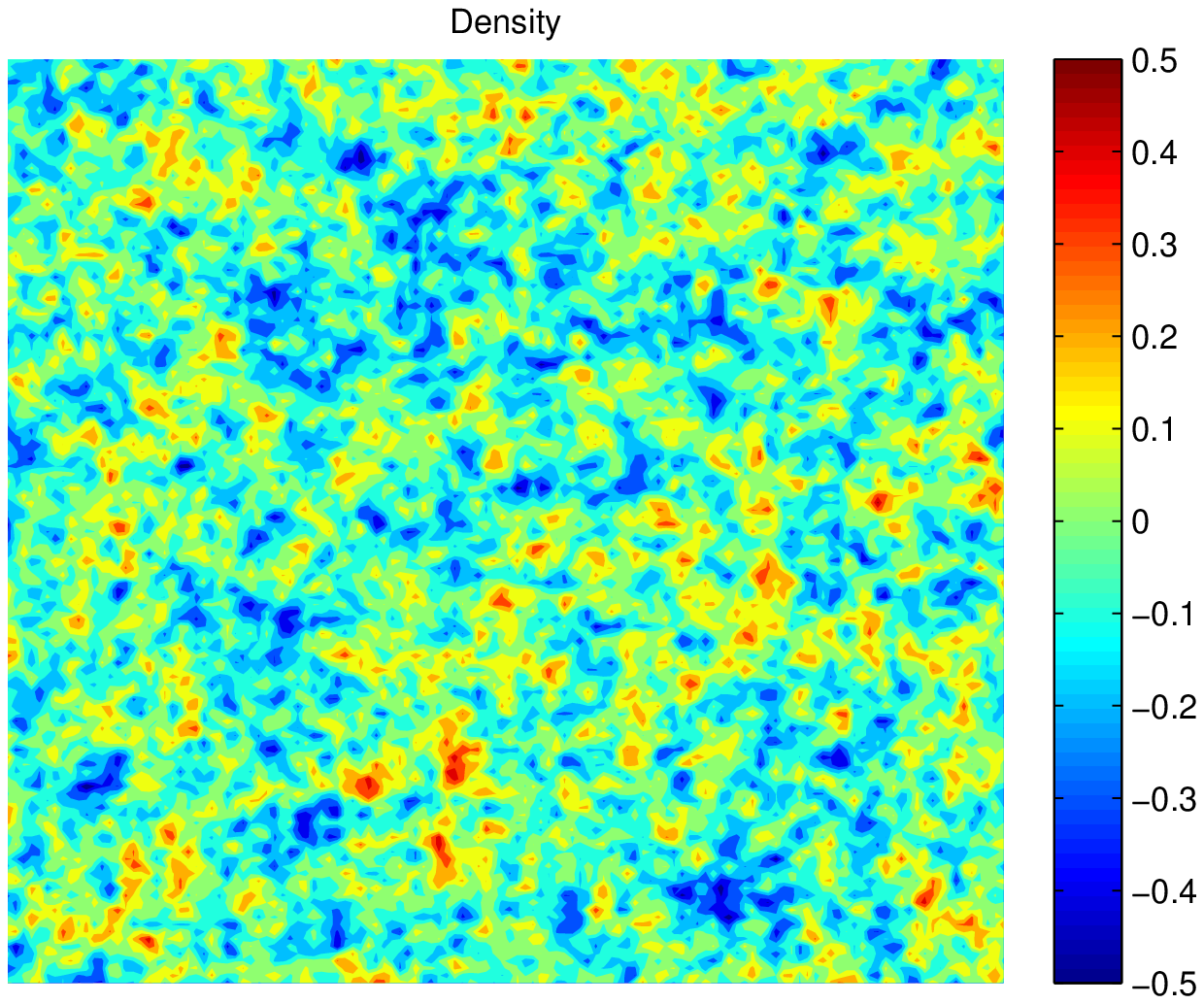}
\includegraphics[width=3.3in]{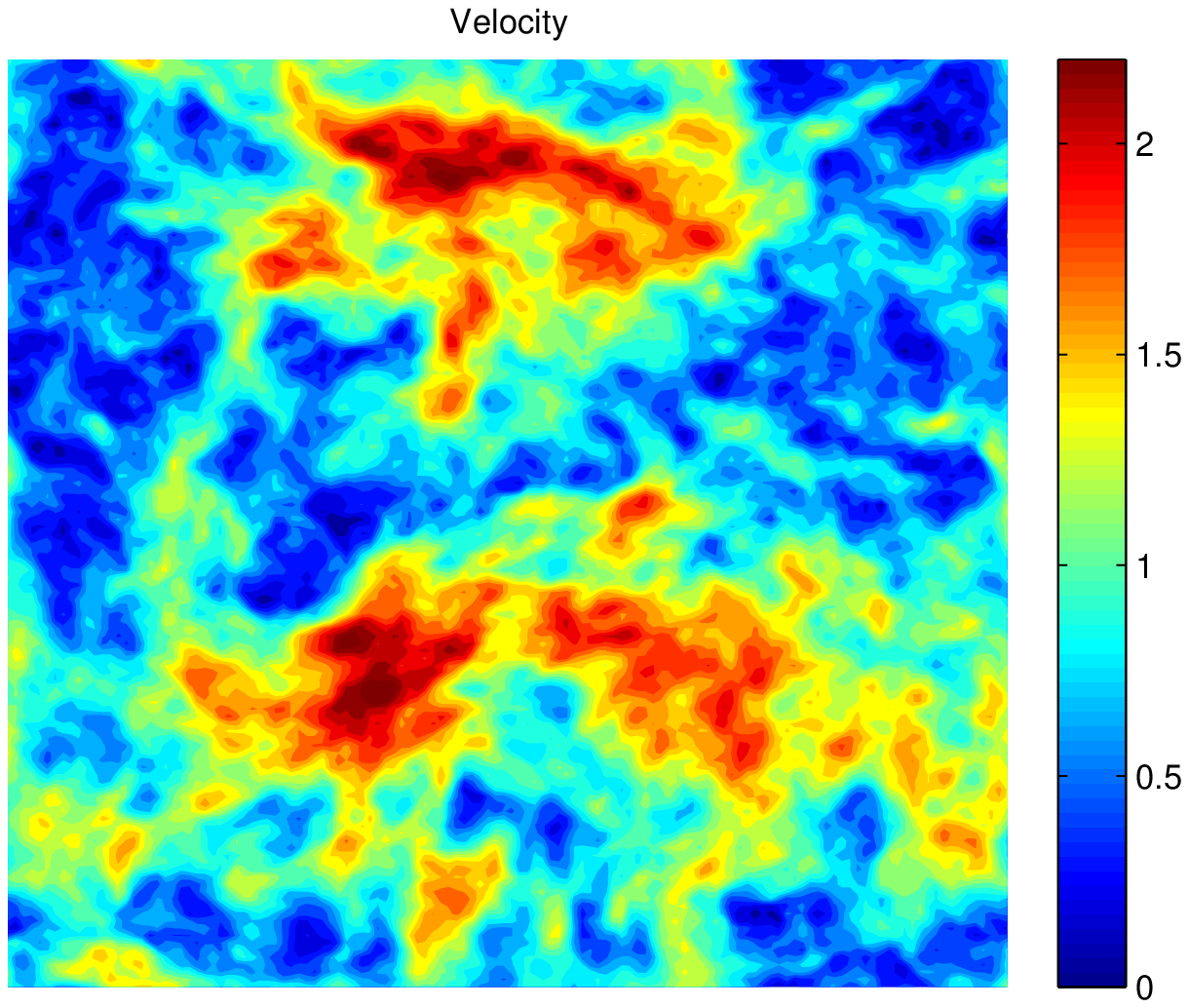}
\caption{The large-scale density and velocity fields in an example of
  a slice from a simulated volume 384 Mpc on a side (based on
  \citet{nature}, but taken from a different box from the one shown in
  the Figures in \citet{nature}, i.e., for a different set of random
  initial conditions). The thickness of the slice is 3 Mpc (which is
  also the pixel size of our grid). For the density field (top panel),
  we show the fractional perturbation relative to the mean, at $z=20$;
  for the velocity field (bottom), we show the magnitude of the
  relative motion in units of the root-mean-square value (the map is
  independent of redshift in these relative units).}
\label{f:RhoV}
\end{figure}

We assumed standard initial perturbations (e.g., from a period of
inflation), where the density and velocity components are Gaussian
random fields. Velocities are coherent on larger scales than the
density, due to the extra factor of $1/k$ in the velocity from the
continuity equation that relates the two fields. This is clearly
apparent in the example shown in Figure~\ref{f:RhoV} of a thin slice
of our simulated volume. The density field fluctuates on relatively
small scales, while the velocity field shows a larger-scale cosmic
web, with coherent structure on scales of order 100~Mpc. This means
that the largest scales, which are easiest to observe, will be
dominated by the pattern due to the velocity effect.

\begin{figure}[]
\includegraphics[width=3.3in]{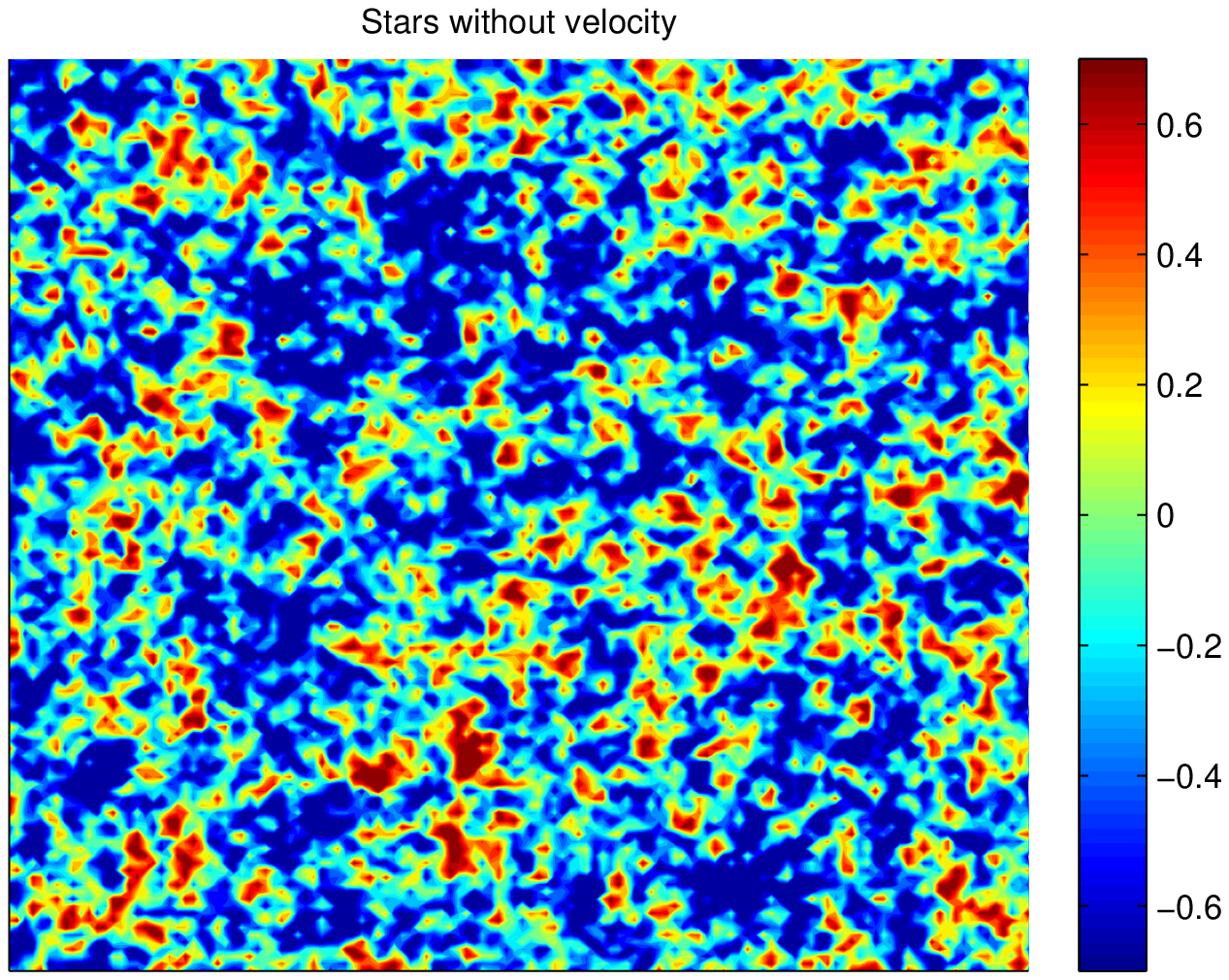}
\includegraphics[width=3.3in]{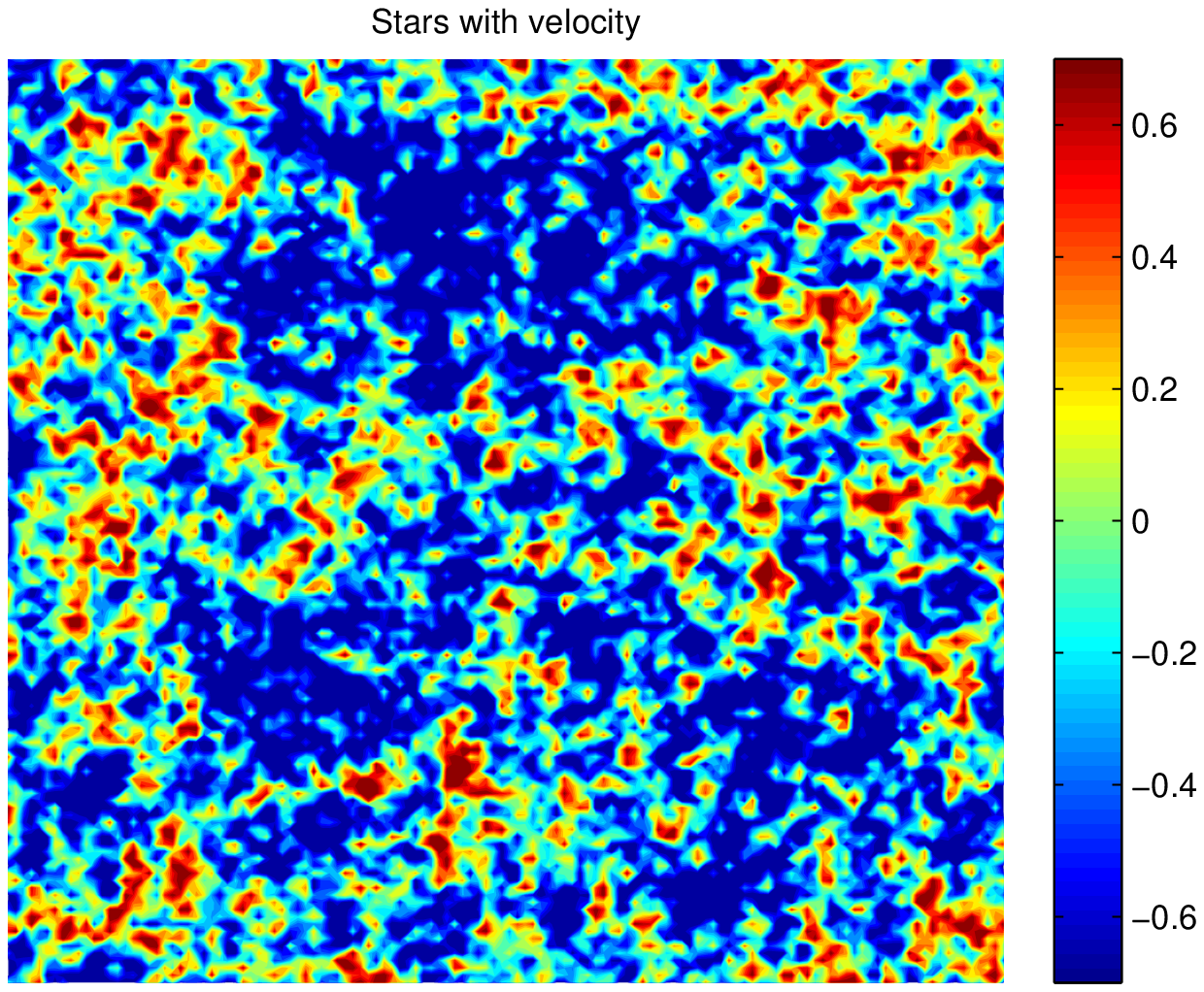}
\caption{Effect of relative velocity on the number density of stars at
  redshift 20. For the same slice as in Figure~\ref{f:RhoV}, we
  compare the previous expectation (top panel), including the effect
  of density only, to the new prediction (bottom), including the
  effect of the same density field plus that of the relative velocity.
  The colors correspond to the logarithm of the gas fraction in units
  of its cosmic mean value in each case.}
\label{f:fgas20}
\end{figure}

\begin{figure}[]
\includegraphics[width=3.3in]{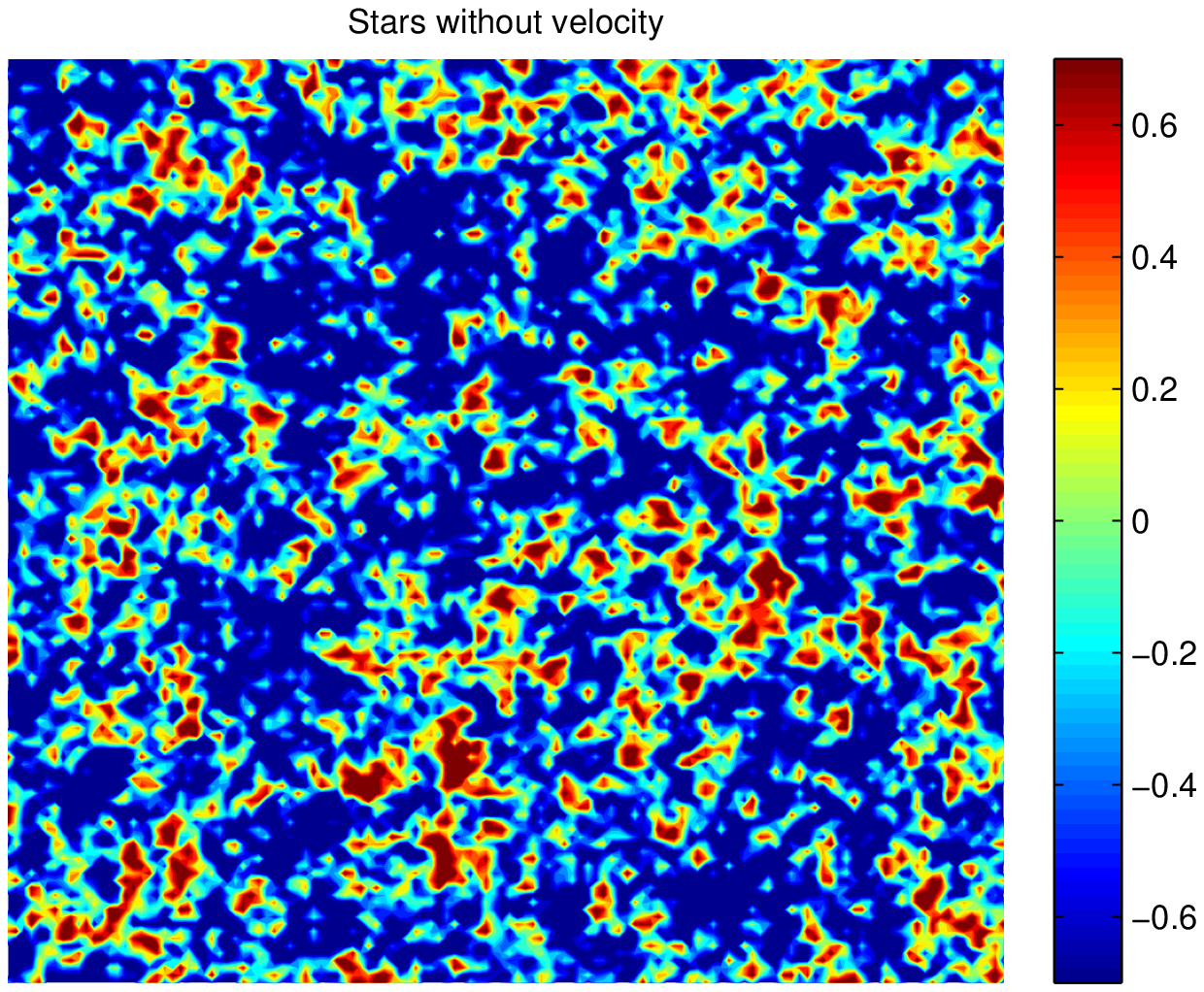}
\includegraphics[width=3.3in]{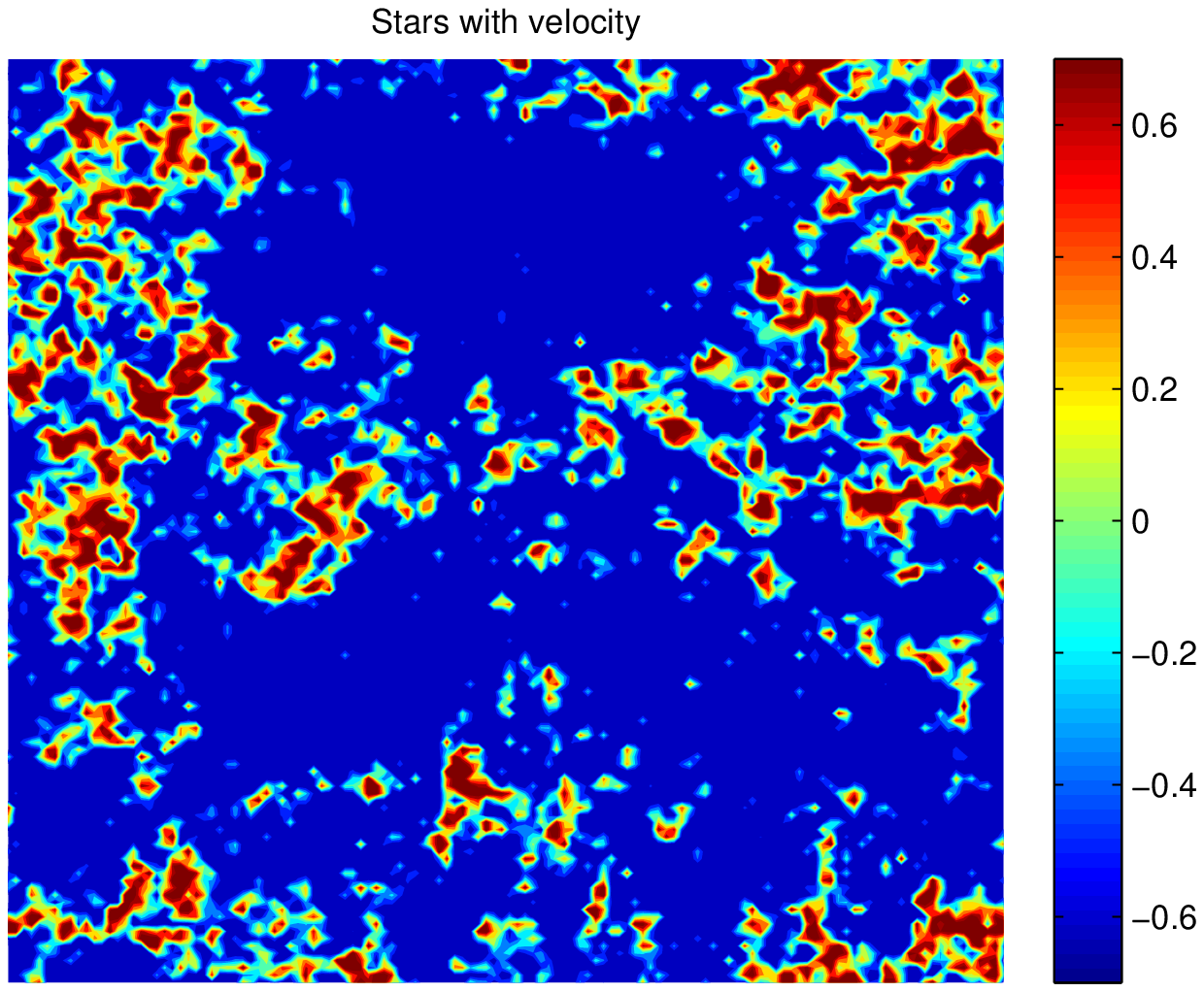}
\caption{Effect of relative velocity on the number density of stars at
  redshift 40. For the same slice as in Figure~\ref{f:RhoV}, we
  compare the previous expectation (top panel), including the effect
  of density only, to the new prediction (bottom), including the
  effect of the same density field plus that of the relative velocity.
  The colors correspond to the logarithm of the gas fraction in units
  of its cosmic mean value in each case. The color scale spans the
  same range as in Figure~\ref{f:fgas20} for easy comparison.}
\label{f:fgas40}
\end{figure}

The resulting distribution of stellar density at $z=20$ is shown in
Figure~\ref{f:fgas20}. Note the large biasing (i.e., amplification of
fluctuations) of the stars: density fluctuations ranging up to $\pm
50\%$ yield (without including the relative velocities) a field of
stellar density that varies by over a factor of 20 (both smoothed on
the 3 Mpc scale). The velocity effect produces a more prominent cosmic
web on large scales, marked by large coherent regions that have a low
density of stars, separated by ribbons or filaments of high star
formation.  The effect is even more striking at higher redshifts
(Figure~\ref{f:fgas40}), which substantially alters the feedback
environment of the first stars. The various types of radiation that
produce feedback spread out to a considerable distance from each
source, but this distance is typically not as large as the span of the
velocity-induced features.  This means that regions of low velocity
(and thus high star formation) experience radiative feedback
substantially earlier than regions of high velocity (low star
formation). Thus, the substantial effect of the velocities on early
star formation makes early feedback far more inhomogeneous than
previously thought.

Observationally, these degree-scale fluctuations will affect various
cosmic radiation backgrounds, and in particular the history of 21-cm
emission and absorption, which depends on the timing of the three
radiative transitions mentioned in Section~\ref{intro21}:
spin-temperature coupling due to \Lya photons, molecular hydrogen
dissociation due to LW photons, and heating due to X-ray photons.
Although the timing is still significantly uncertain, the 21-cm
coupling due to \Lya radiation is expected to occur rather early, with
the X-ray heating fluctuations occurring later and possibly
overlapping with significant small-halo suppression due to LW
radiation \citep{nature}. It is most promising to focus on the
fluctuations due to X-ray heating around redshift 20, since these tend
to produce larger 21-cm fluctuations (and at lower redshifts that
feature lower foreground noise) than the \Lya coupling era [as
suggested by previous analyses of these two periods without the
relative velocity effect \citep{BL05b,Jonathan07}]. Given the expected
timing, we can analyze the heating era assuming that \Lya coupling has
already saturated. As for the LW flux, we continue to illustrate here
the case of negligible LW feedback (as was assumed in
Figures~\ref{f:fgas20} and \ref{f:fgas40}), but below we bracket the
effect of the LW flux by also considering the opposite limiting case
where the LW transition has already saturated.

\begin{figure}[]
\includegraphics[width=3.3in]{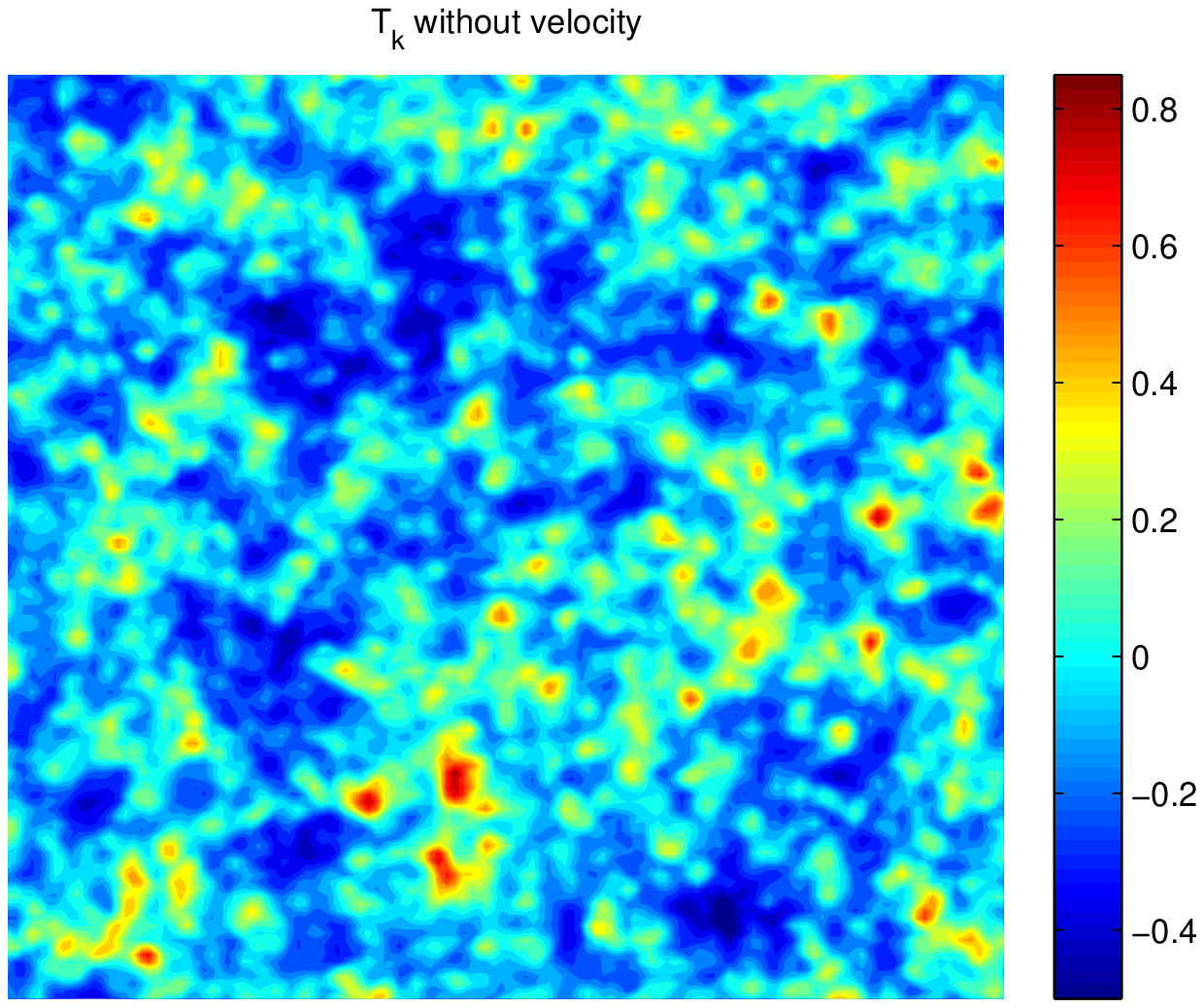}
\includegraphics[width=3.3in]{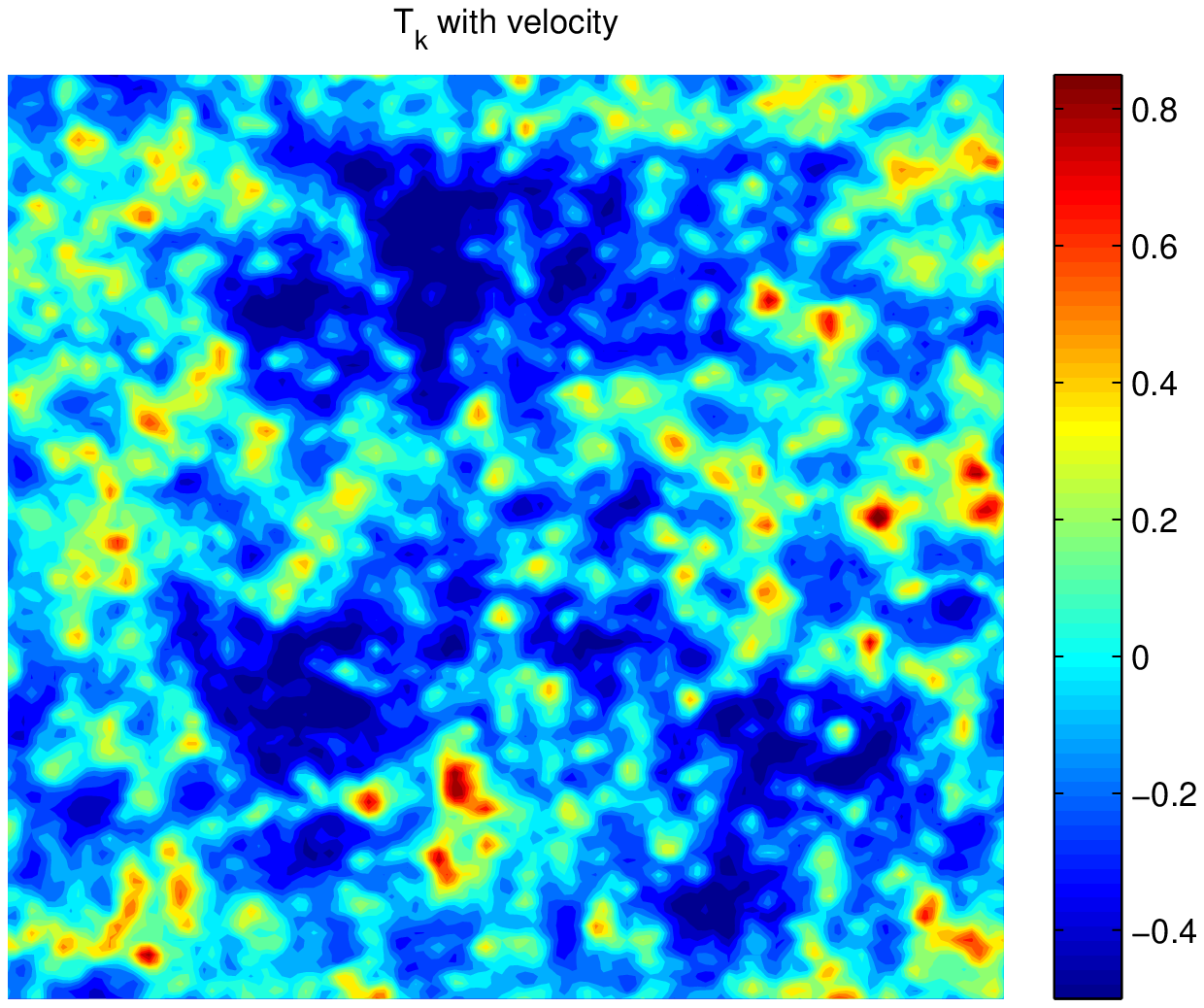}
\caption{Effect of relative velocity on the gas temperature $T_k$ at
  redshift 20. For the same slice as in Figure~\ref{f:RhoV}, we
  compare the previous expectation (top panel), including the effect
  of density only, to the new prediction (bottom), including the
  effect of density and relative velocity. The colors correspond to
  the logarithm of the gas (kinetic) temperature in units of the CMB
  temperature at $z=20$.}
\label{f:Tk20}
\end{figure}

Figure~\ref{f:Tk20} shows the gas temperature distribution at $z=20$,
at the midpoint of the heating transition, i.e., when the mean gas
temperature is equal to that of the CMB. Regions where the gas moved
rapidly with respect to the dark matter (dark red regions, bottom
panel of Figure~\ref{f:RhoV}) produced fewer stars (dark blue regions,
bottom panel of Figure~\ref{f:fgas20}) and thus a lower X-ray
intensity, leaving large regions with gas that is still colder than
the CMB by a factor of several (dark blue regions, bottom panel of
Figure~\ref{f:Tk20}). The spatial reach of X-rays results in a gas
temperature distribution that is smoother than the the distribution of
stars, and this brings out the effect of large-scale fluctuations and
thus highlights the contrast between the effect of density and
velocity fluctuations.

\begin{figure}[]
\includegraphics[width=3.3in]{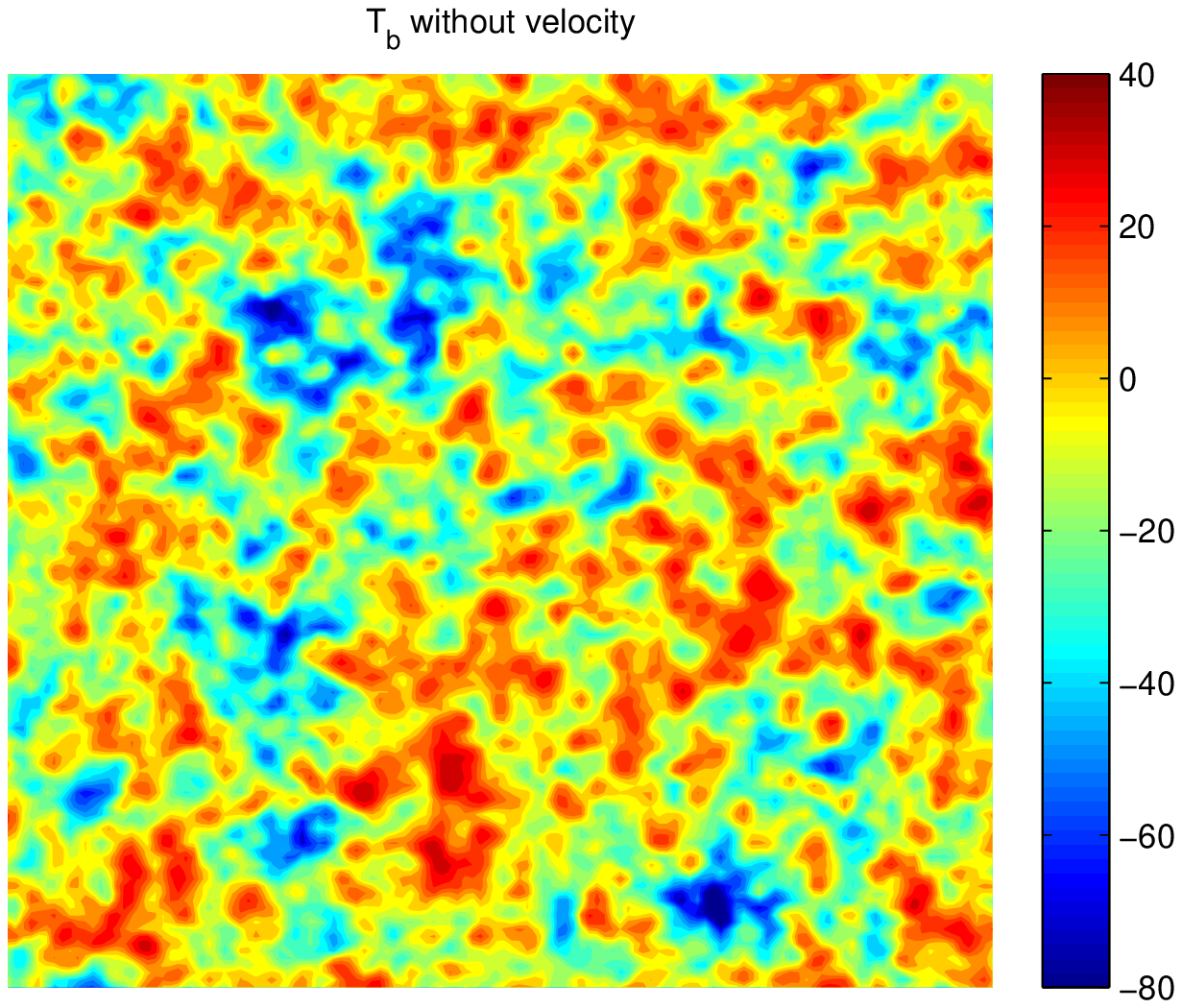}
\includegraphics[width=3.3in]{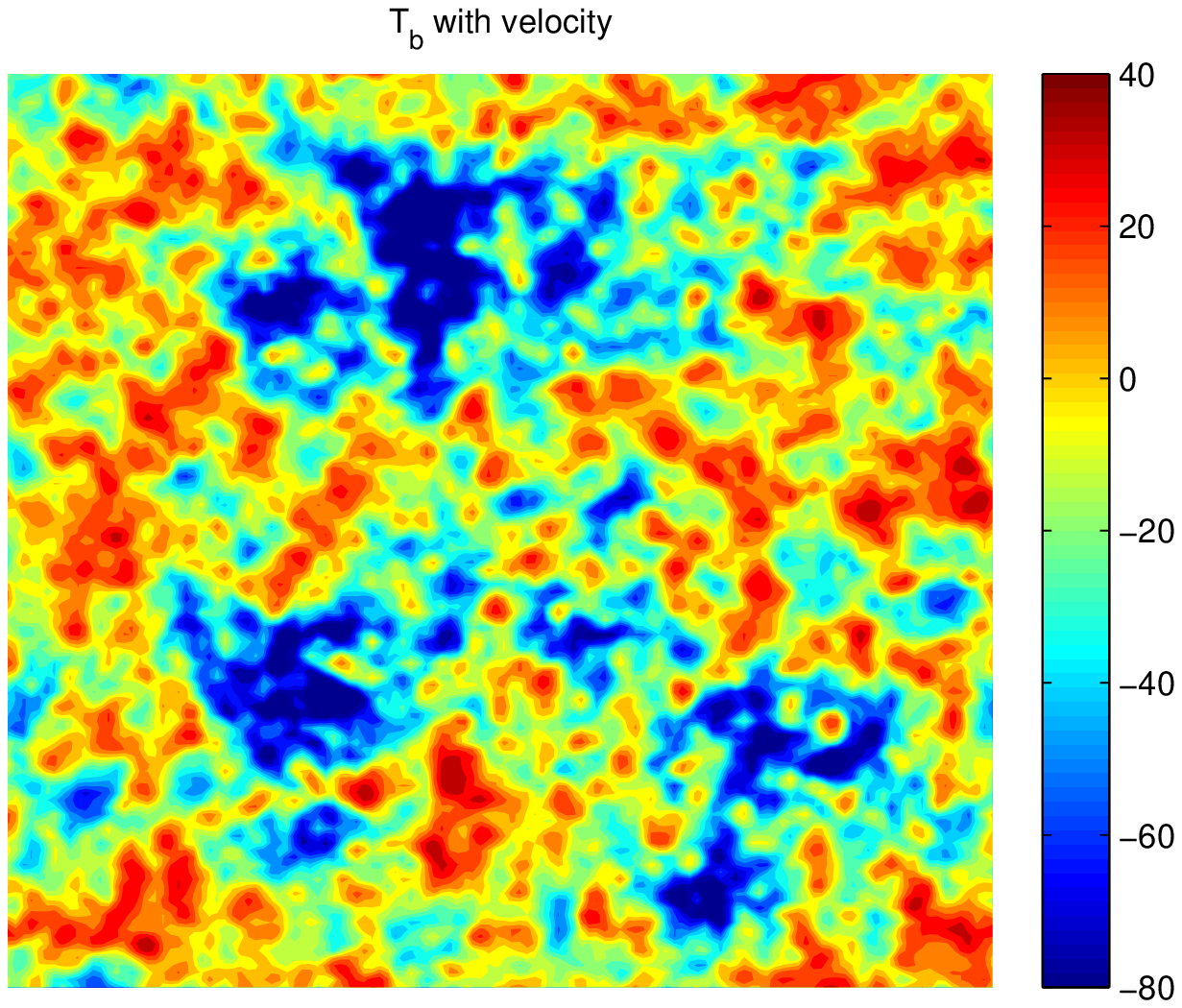}
\caption{Effect of relative velocity on the redshift 20 21-cm
  brightness temperature $T_b$ (which measures the observed intensity
  of radio waves emitted by intergalactic hydrogen atoms at 21~cm).
  For the same slice as in Figure~\ref{f:RhoV}, we compare the
  previous expectation (top panel), including the effect of density
  only to the new prediction (bottom), including the effect of density
  and relative velocity. The colors correspond to the 21-cm brightness
  temperature in millikelvin units.}
\label{f:Tb20}
\end{figure}

During the heating transition, the 21-cm brightness temperature
(Figure~\ref{f:Tb20}) mainly measures the gas temperature, although it
is also proportional to the gas density (and to the square root of
$1+z$). The form of the dependence, $T_{\rm b} \propto 1 - T_{\rm
CMB}/T_{\rm gas}$, makes the 21-cm intensity more sensitive to cold
gas than to hot gas (relative to the CMB temperature). Thus, the large
voids in star formation produced by a high relative velocity lead to
prominent 21-cm absorption (dark blue regions, bottom panel of
Figure~\ref{f:Tb20}) seen on top of the pattern from the effect of
density fluctuations. These deep 21-cm cold spots are the main
observable signature of the effect of the relative velocity on the
first stars. Note that the observed wavelength of this radiation is
redshifted by the expansion of the universe to 4.4 meters
(corresponding to a frequency of 68 MHz).

\begin{figure}[]
\includegraphics[width=3.3in]{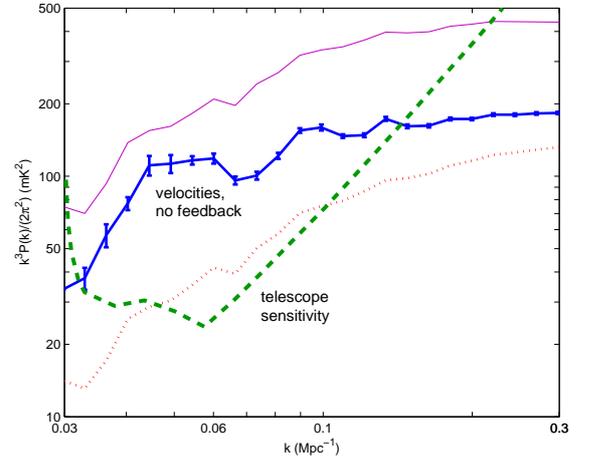}
\caption{Signature of the relative velocity in the 21-cm power
  spectrum, at the peak of the X-ray heating transition at $z=20$. We
  consider the prediction including the relative velocity effect (blue
  solid curve) or with the effect of densities only (red dotted
  curve), both for the case of a late LW transition for which the LW
  feedback is still negligible at $z=20$. These predictions are
  compared to the projected 1-$\sigma$ telescope sensitivity (green
  dashed curve) based on 1000-hour observations with an instrument
  like the Murchison Wide-field Array or the Low Frequency Array but
  designed to operate at 50--100 MHz \citep{McQuinn}, where we include
  an estimated degradation factor due to foreground removal
  \citep{Liu}; this sensitivity is defined as the signal that would
  yield a measurement with a signal-to-noise ratio of unity in each
  $k$ bin of size $\Delta k = 0.5 k$ averaged over an 8~MHz frequency
  band. Future experiments like the Square Kilometer Array should
  reach a better sensitivity by more than an order of magnitude
  \citep{McQuinn}. To allow for the possibility of feedback, we also
  show the prediction for the opposite limit of maximum feedback,
  i.e., an early LW transition that has already saturated (purple
  solid curve). In this plot we have fixed the heating transition at
  $z=20$ for easy comparison among the various cases. From
  \citet{nature}.}
\label{f:nPS}
\end{figure}

While these Figures illustrate the detailed pattern of the effect of
relative velocity on the 21-cm intensity distribution, upcoming
experiments are expected to yield very noisy maps that likely must be
analyzed statistically. Figure~\ref{f:nPS} shows the predicted effect
on a key statistic, the power spectrum of the fluctuations in 21-cm
intensity (from \citet{nature}). The velocities enhance large-scale
fluctuations (blue solid curve compared with red dotted), leading to a
flatter power spectrum with prominent baryon acoustic oscillations
(reflecting the BAO signature in Figure~\ref{f:vbc}). The signal is
potentially observable with a redshift 20 version of current
instruments (green dashed curve). If there is strong LW feedback
(solid purple curve), then the small galaxies that rely on
molecular-hydrogen cooling are unable to form; the larger galaxies
that dominate in that case are almost unaffected by the streaming
velocity, so the 21-cm power spectrum reverts to the density-dominated
shape (compare the solid purple and red dotted curves), but it becomes
even higher since more massive galactic halos are even more strongly
biased (i.e., clustered).

Thus, regardless of the timing of the LW feedback, the 21-cm power
spectrum at the peak of the heating transition should feature large
observable fluctuations on large angular scales. Beyond just detection
of the signal, only a mild additional accuracy is necessary in order
to determine whether feedback has suppressed star formation in the
smallest halos.  If it has not, than the velocity effect produces
strong BAOs on top of a flattened power spectrum, in particular
raising it by a factor of 4 on large scales ($k=0.05$ Mpc$^{-1}$,
wavelength 130~Mpc, observed angle 2/3 of a degree) where experimental
sensitivity is optimal.  If this characteristic shape is observed it
would confirm that million mass halos dominated galaxy formation at
this early epoch.

In summary, the velocity effect makes it significantly easier to
detect the 21-cm power spectrum during the epoch of the first heating
of the IGM. It also creates a clear signature by flattening the power
spectrum and increasing the prominence of the BAOs (which are more
strongly imprinted in the velocity than in the density fluctuations).
While Figure~\ref{f:nPS} considers a single redshift, similar
observations over the full $\Delta z \sim 6$ redshift range of
significant heating fluctuations could actually detect the slow
advance of the LW feedback process, during which the power spectrum is
predicted to continuously change shape, gradually steepening as the
BAO signature weakens towards low redshift.

\section{Open Questions}

As summarized above, the recently-noticed velocity streaming effect
causes a special form of luminosity bias at high redshift. Unlike the
standard galaxy bias relative to the underlying density field, early
galaxies were subject to a bias that depends on the velocity field.
Specifically, the velocity difference between the baryons and dark
matter affected the smallest star-forming halos in which the gas
cooled via molecular hydrogen cooling. This bias of luminosity
relative to velocity results in large fluctuations in the abundance of
the first stars. Since the velocity varies on larger scales than the
density field, and larger scales are easier to observe (since lower
resolution suffices to detect the corresponding 21-cm fluctuations),
the velocity effect immensely increases the feasibility for upcoming
21-cm observations to detect stars from higher redshifts ($\sim
15-25$) than previously expected.  Moreover, such a detection would be
both easier to recognize and more exciting than previously thought.

These developments should make the whole pre-reionization epoch far
more attractive and interesting to the whole community of theorists,
numerical simulators, and observers. In fact, this era is rich with
little-explored astrophysics. As the first stars formed, their
radiation (plus emission from stellar remnants) produced feedback that
radically affected not only the IGM but also the character of
newly-forming stars. As shown above, the velocity effect substantially
modifies the distribution of the first stellar generations, and in
fact the star-formation history is changed further since radiative
feedback (which affects all stars beyond the very first ones) itself
depends on this distribution. 

In the previous section we have mentioned only some first explorations
of the radiative feedback during this early era. However, in order to
more reliably guide observers to the expected 21-cm power spectrum
from this era, the radiative feedback should be explored in much
greater detail. For example, the 21-cm fluctuations generated during
the heating transition, studied by  \citet{Jonathan07} without the
streaming velocity and by  \citet{nature} with the velocity included,
should be studied together with inhomogeneous LW feedback. Also, the
21-cm fluctuations during \Lya coupling originally studied by
 \citet{BL05b} should be studied in the presence of the streaming
velocity and the LW feedback. A full study of these potentially
observable signals will also require a new generation of numerical
simulations that will explore the small-scale interaction between
relative velocity and star formation in the presence of various
feedbacks such as heating and LW feedback.

At the same time, significant uncertainty is likely to always remain
in any prediction that involves complex astrophysics such as star
formation and feedback. Thus, only observations can ultimately confirm
or exclude the promising theoretical predictions shown in the previous
section. We therefore expect increased observational efforts focused
on this early epoch. Such observations would push well past the
current frontier of cosmic reionization ($z \sim 10$, $t \sim
480$~Myr) for galaxy searches and 21-cm arrays, and thus would
represent a new frontier for extragalactic astronomy.

\section*{Acknowledgments} 

I thank my collaborators on the reviewed work, especially Anastasia
Fialkov, Eli Visbal, Dmitriy Tseliakhovich, Chris Hirata, Avi Loeb,
and Smadar Naoz. This work was supported by Israel Science Foundation
grant 823/09.


\end{document}